# Undulating Diamond Photonic Crystal in the Royal Jewel Butterfly: Circumventing Refractive Index Limits to Defy Iridescence

*Yuanbo Li,*[1] *Shuo Huang,*[2] *Xi Wang,*[1] *Congcong Cui,*[1] *Hao Chen,*[3] *Yuqin Xiong,*[4] *Wang Zhang,*[4] *Matthias Saba,*[5,6,*] *Xueyan Feng,*[2,*] *Bo Chen,*[7,8,*] *Yuanyuan Cao,*[9,*] *& Lu Han*[1,*]

[1] School of Chemical Science and Engineering, Tongji University; Shanghai 200092, China. [2] State Key Laboratory of Molecular Engineering of Polymers, Department of Macromolecular Science, Fudan University; Shanghai 200438, China. [3] Institute of Mathematical Sciences, ShanghaiTech University; 393 Middle Huaxia Road, Shanghai 201210, China. [4] State Key Laboratory of Metal Matrix Composite, School of Materials Science and Engineering, Shanghai Jiao Tong University; Shanghai 200240, China. [5] Adolphe Merkle Institute, University of Fribourg; Fribourg, Switzerland. [6] NCCR Bio-inspired Materials, University of Fribourg; Fribourg, Switzerland. [7] Key Laboratory of Advanced Civil Engineering Materials of the Ministry of Education, School of Materials Science and Engineering, Tongji University; Shanghai 201804, China. [8] Hubei Key Laboratory of Biomass Fibers and Eco-dyeing & Finishing, School of Textile Science and Engineering, Wuhan Textile University; Wuhan 430200, China. [9] Laboratory of Low-Dimensional Materials Chemistry, Key Laboratory for Ultrafine Materials of Ministry of Education, School of Materials Science and Engineering, East China University of Science and Technology; Shanghai 200237, China. *e-mails: matthias.saba@unifr.ch; fengxueyan@fudan.edu.cn; bo.chen@tongji.edu.cn; yuanyuancao@ecust.edu.cn; luhan@tongji.edu.cn

**The modest refractive-indices of biomolecules place fundamental constraints on the optical performance of natural photonic organisms. To overcome this limitation, organisms reshape basic optical structures into hierarchical architectures, exhibiting performance rivals that of high-index materials. Here we show how the Royal Jewel butterfly (*Hypochrysops polycletus*) employs such a strategy to circumvent the low refractive-index of chitin to achieve an unusual combination of intense structural colouration and remarkably weak angular dependence. A single-diamond photonic crystal, which was hailed as the "holy grail" and documented only in beetles and weevils, has now been discovered in the butterfly. It deforms into a hierarchically undulated superstructure guided by basal lamina and correlated with quasi-twinning boundaries with mirror symmetry. The scale surface is thus fully terminated by inclined {111} facets essential for fundamental bandgap, producing bright specular Bragg reflection at intermediate to high angles and low-angle radiation enabled by simultaneously coupling localised resonances populated through anisotropic Bloch transport, thereby transforming an otherwise strongly angle-dependent photonic crystal into a vivid, angle-robust optical structure. Following this optical concept, an artificially angle-insensitive optical prototype was fabricated through two-photon polymerisation, validating the mechanism as a transferable design principle at the intersection of fundamental nanophotonics and bio-inspired engineering.**

High refractive-index ($n$) contrast is essential for various photonic devices, with prominent examples including photonic crystals, multilayer reflectors, high-index-contrast waveguides and dielectric metasurfaces[1-5]. Biological photonic structures[6,7], however, are constructed from a limited palette of biopolymers such as chitin, cellulose and proteins, which generally possess relatively modest refractive-indices in the visible range[8-13]. This refractive-index constraint places fundamental limits on optical performance and requires biological systems to rely on structural organisation rather than material composition to achieve advanced photonic functionality[14,15].

Three-dimensional (3D) photonic crystals (PCs) provide a particularly powerful optical platform because their periodic nanoscale architecture controls optical propagation in all spatial directions[16,17]. In nature, these structures often adopt bicontinuous, triply periodic morphologies bounded by saddle-shaped surfaces with negative Gaussian curvature. Prominent examples include the single-gyroid (SG), commonly found in butterfly wing scales[18,19], and the single-diamond (SD), previously reported only in beetles, most notably in weevils[12,13,20-22]. The SD topology is particularly attractive because, at sufficiently high refractive-index contrast ($n > 1.9$), it supports one of the widest known complete photonic bandgaps, considered the "holy grail" of PC[23-25]. However, the refractive-indices of various biological materials (e.g., chitin, $n \approx 1.55$) lie far below this high-contrast regime, such that their 3D photonic crystals operate via directional partial stopbands whose optical response strongly depends on crystal orientation and viewing direction. Commonly, this angular-dependent iridescence could be reduced through polycrystallinity or correlated disorder; however, the resulting angular uniformity comes at the expense of reflectivity and colour saturation[18,26-28].

On the other hand, nature has evolved hierarchical integration of optical structures across multiple length scales to overcome refractive-index limitations, allowing larger-scale morphology to reshape, concentrate, or redistribute the response of nanoscale photonic elements. Examples include the narrow, intense reflections through multilayer structures embedded in concavities, and strong light-dispersion efficiency through diffraction gratings arranged on curved or oppositely inclined surfaces[29-34]; hierarchical spherical or lamellar structures assembled from biogenic guanine and isoxanthopterin nanoplates that show optical functions comparable to $TiO_2$- and ZnS-based materials[35-38]. These strategies provide examples that simultaneously satisfy competing demands on optical performance, material availability, formability and mechanical robustness[6], and thus offer valuable starting points for designing optical materials in otherwise high-dimensional parameter spaces[23,25].

Here, we show how the Royal Jewel butterfly, *Hypochrysops polycletus* (*H. polycletus*) uses the structure deformation strategy to circumvent the low refractive index of chitin, achieving strongly brilliant and uniformly blue angle-independent structural colour through the hierarchical organisation of the SD 3D PC. This Lycaenidae butterfly, distributed across northeastern Australia, eastern Indonesia and Papua New Guinea, is renowned for its sapphire-like appearance[33,39,40]. Using advanced 3D electron microscopy, angle-resolved optical measurements and electromagnetic simulations, we identify the SD photonic crystal within the scale surface, extending this topology to Lepidoptera for the first time[19,41,42]. Remarkably, guided by the wavy morphology of the scale basal lamina, the SD network is integrated into a quasi-sinusoidal superstructure composed of wavy, quasi-twinned crystals that exhibit pronounced curvature at the crests and gentle curvature with continued growth in the valleys, fully exposing the crucial {111} facets towards the scale surface with a series of inclined angles. Those hierarchical structures could create different specular mechanisms at different viewing directions, complementarily yielding a high degree of angular robustness even under different orientations and structural perturbations. Consequently, this design resolves the long-standing trade-off between bright-but-iridescent periodic PCs and dim-yet-angle-robust polycrystalline or correlated-disordered structures, offering a practical blueprint for future nanophotonics.

## Appearance and optical properties of wing scales of the Royal Jewel butterfly

Unlike other species with iridescence or muted hues, the Royal Jewel butterfly shows an almost angle-independent specular bright blue lustre, which remains constant when viewed along both the longitudinal and transverse directions of the scale (Fig. 1a and 1b, Supplementary Fig. 1 and Supplementary Video 1).

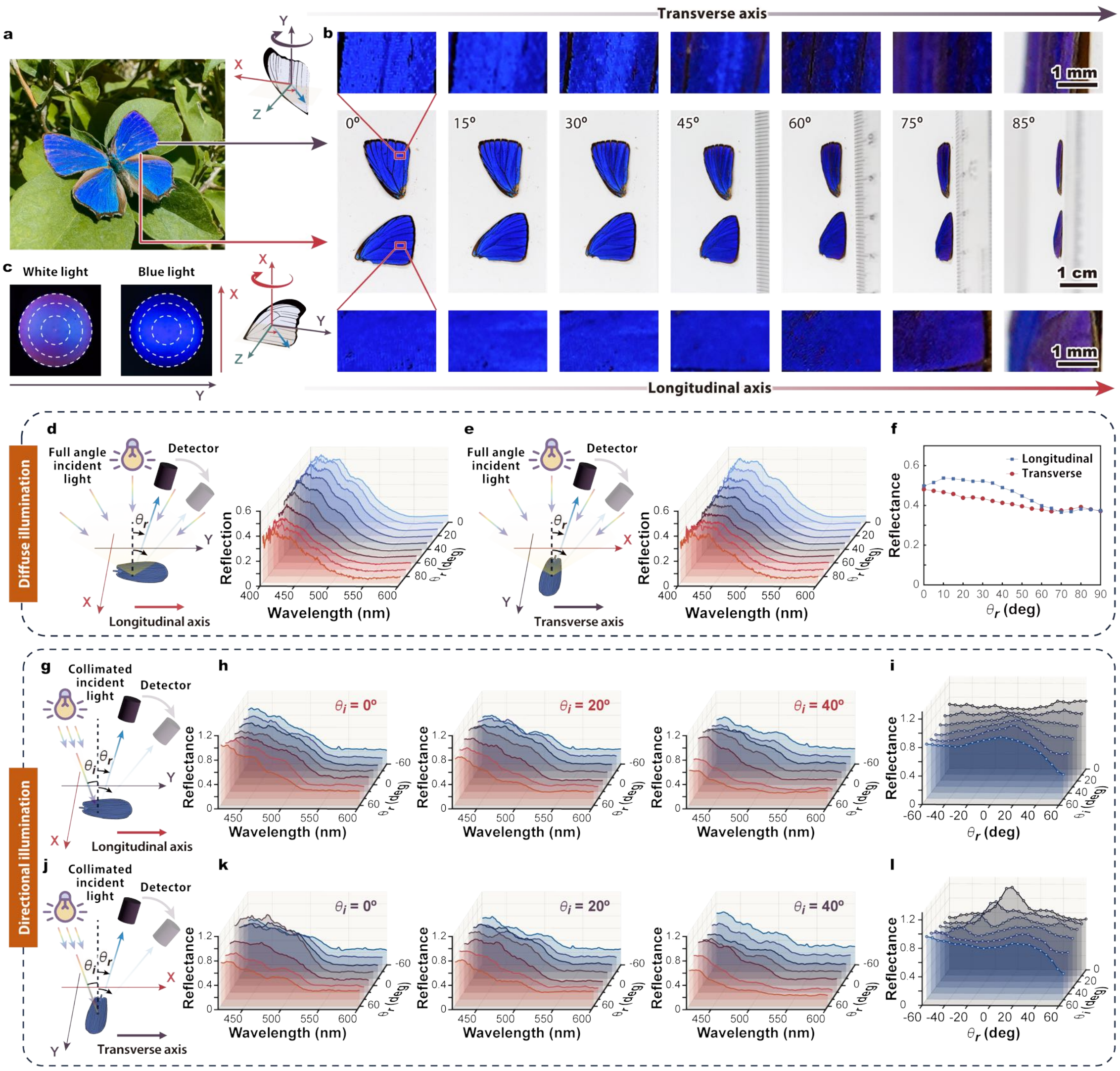

**Fig. 1. Appearance and optical properties of the Royal Jewel butterfly. a,** Optical photograph of the *H. polycletus* butterfly illuminated by natural sunlight. **b,** Appearance of a dorsal wing of the butterfly, viewed after tilting the wing along its longitudinal (bottom) and transverse (top) axis. **c,** Back-focal-plane images of a scale under white light (left) and blue light (right). Circles indicate scattering angles of 20°, 40° and 60°. **d,e,** Schematic of the ARMS (left) of the dorsal wing surface under diffuse illumination with variable detection angles along the longitudinal (**d**) and transverse axis (**e**) and the corresponding angle-resolved reflection spectra (right). **f,** Plot of average reflectance over a 418-435 nm wavelength range viewed from different angles. **g,** Schematic drawing of the ARMS of the dorsal wing surface using directional illumination with the detector position changed along the longitudinal axis. **h,** ARMS of dorsal wing surface under different incident angles. **i,** Plot of average reflectance over the 418-435 nm wavelength range from different incident and reflection angles. **j,** Schematic of the ARMS of the dorsal wing surface using directional illumination under different incident angles along the transverse axis. **k,** Corresponding ARMS of dorsal wing surface under different incident angles. **l,** Plot of average reflectance over the 418-435 nm wavelength ranges from different incident and reflection angles.

The back-focal-plane image that maps *k*-space also confirms this homogeneous blue hue, particularly along the longitudinal axis for scattering angles up to 60°, which is the limit set by the numerical aperture and the maximum incident angle (Fig. 1c). To quantitatively characterise this unique optical behaviour, we performed two types of angle-resolved micro-spectroscopy (ARMS) measurements on the dorsal wing:

*1) Diffuse illumination measurement.* The incident light spanning polar angles between ±60° was focused on a single scale to emulate natural lighting conditions, and the reflected intensity was recorded at different polar angles ($\theta_r$) along the longitudinal and transverse directions of the wing (Fig. 1d-f and Supplementary Fig. 2). Polytetrafluoroethylene (PTFE) was used as a Lambertian reference for normalisation. A broad spectral response below 500 nm was observed, with the peak position remaining essentially unchanged across a wide angular range (Fig. 1d,e). The averaged relative reflectance in the 418-435 nm range confirmed that the reflectance remains relatively high, particularly along the longitudinal axis, while the transverse axis shows a gentle linear attenuation (Fig. 1f and Supplementary Fig. 2).

*2) Directional illumination.* A collimated light source was employed to probe the relationship between the illumination and reflection angles (Fig. 1g,j, Supplementary Fig. 3, and Supplementary Fig. 4). The incident angle $\theta_i$ varied from 0 to 50°, while reflection spectra were recorded from –60° to +60°. The incident wavelengths spanned 418-693 nm, matching the halogen lamp's 418-nm lower limit and covering the reflection range (Supplementary Fig. 5a). The spectra from the longitudinal axis exhibit excellent angular stability, with only a slight drop at large incident and reflection angles caused by the Fresnel effect of the PTFE reference (Fig. 1g,h). In contrast, although reflectance in the spectra from the transverse axis remains relatively high overall, a more pronounced intensity decreasing with detection angle has been observed (Fig. 1j,k). Mean reflectance within the 418-435 nm spectral band mirrors this trend (Fig. 1i,l). Normalised intensity thus provides an experimental proxy for the bidirectional reflectance distribution

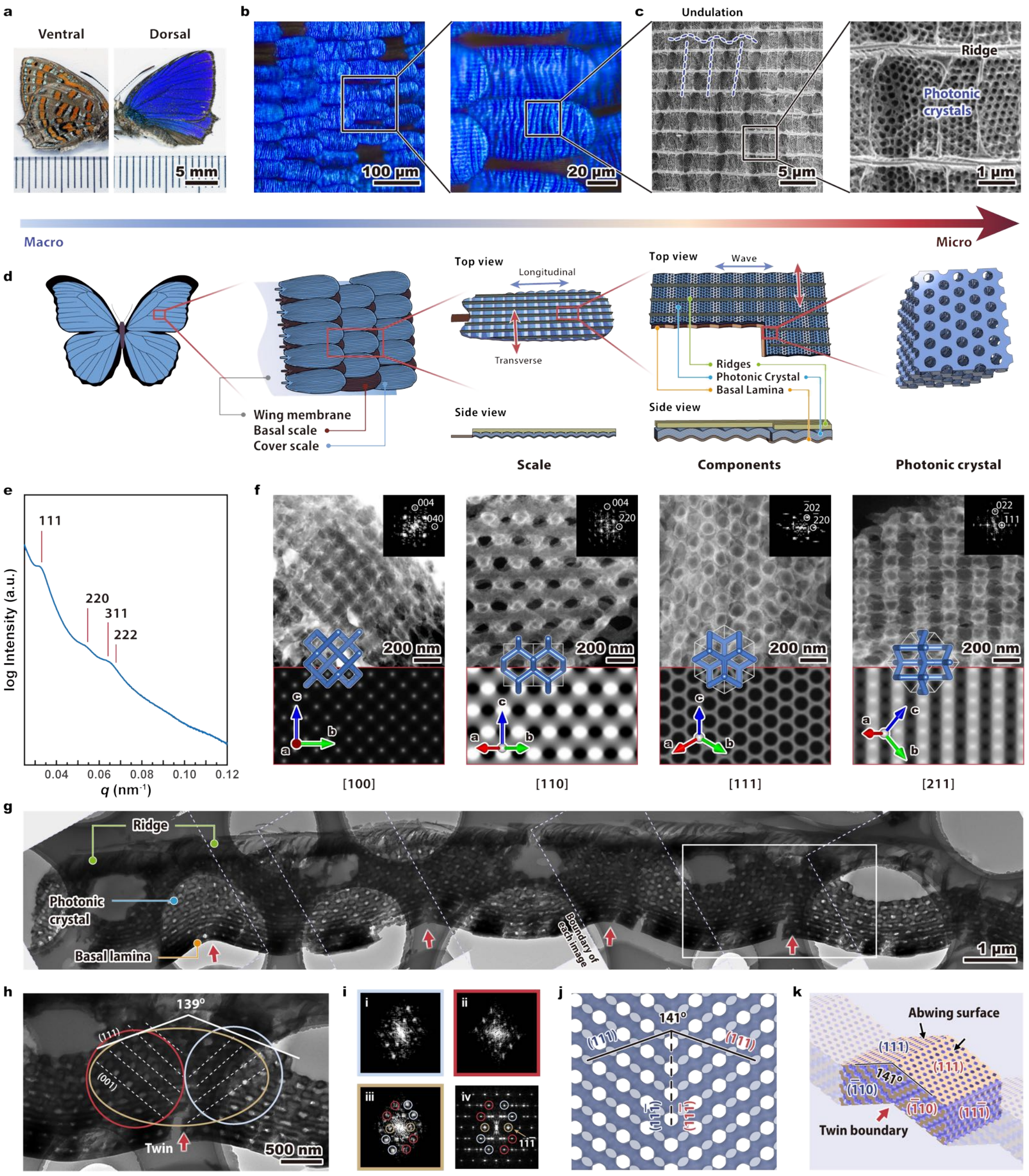

**Fig. 2. Structural investigations of dorsal wing scales of the Royal Jewel butterfly. a,** Optical photograph of the dorsal (right) and ventral (left) surface of *H. polycletus* wings. **b,** Low magnification (left) and high magnification (right) visible light microscopy images of the dorsal surface. **c,** SEM images of the dorsal scales showing the PC structure and ridges. **d,** Schematic diagram of the hierarchical photonic system on the Royal Jewel butterfly wing scales. **e,** SAXS pattern measured from the wing. **f,** HAADF-STEM images and the corresponding Fourier diagrams (FDs) of the photonic structure on the wing scale taken from [100], [110], [111] and [211], respectively. The corresponding theoretical images of SD, generated with SPIRE[45], are shown below the images, with stick models superimposed on both the TEM and theoretical images. **g,** Mosaicked TEM image composed of seven sub-images. The boundaries of the structural domains are evident at the crests and valleys. **h-k,** Boxed area from **g** showing the SD twinning (**h**) and the corresponding FDs taken from twin individuals and the whole region (**i**). **j,k,** Schematic models show that the twin individuals contact each other at the consecutive $(11\bar{1})$ plane running vertically, exposing $\{111\}$ facets that form a quasi-sinusoidal abwing surface.

function (BRDF) of the sample.

Additionally, to rule out pigments' contributions to the chromaticity, refractive index matching experiments were performed, confirming the structural origin of the blue colouration (Supplementary Note 1 and Supplementary Fig. 5).

## Structure determination of the wing scales

To uncover the underlying optical phenomena, we investigated the structural origin of the wing. The dorsal wings are covered by elongated, brilliant blue scales aligned along the longitudinal axis (Fig. 2a). Optical microscopy revealed that the scales consist of numerous transverse blue stripes (~4 µm wide) perpendicular to the longitudinal axis (Fig. 2b). Unlike *C. rubi* and *T. opisena*[18,19], no mosaic of crystalline domains was found. Scanning electron microscopy (SEM) reveals undulating transverse structures with an imperfect ~3-4 µm periodicity that matches the stripe pattern beneath the longitudinal ridges (Fig. 2c, Supplementary Fig. 6). These transverse structures consist of a periodic 3D skeletal scaffold characteristic of a PC with hexagonal-like surface pores (Fig. 2c right and Supplementary Fig. 7). Fig. 2d illustrates this structural hierarchy, progressing from the wing through individual scales down to the nanoscale undulating microstructure that underpins the PC lattice.

To resolve the PC structure, the wings were analysed through small-angle X-ray scattering (SAXS) measurements (Fig. 2e), which suggest a face-centred cubic lattice with an effective lattice parameter of ~325 nm (Supplementary Note 2). The sample was subjected to thin sectioning for transmission electron microscopy (TEM) investigations. High-angle annular dark-field scanning TEM (HAADF-STEM) images and the corresponding Fourier diffractograms (FDs) show both four-fold and six-fold symmetries, typical for a cubic unit cell (Fig.

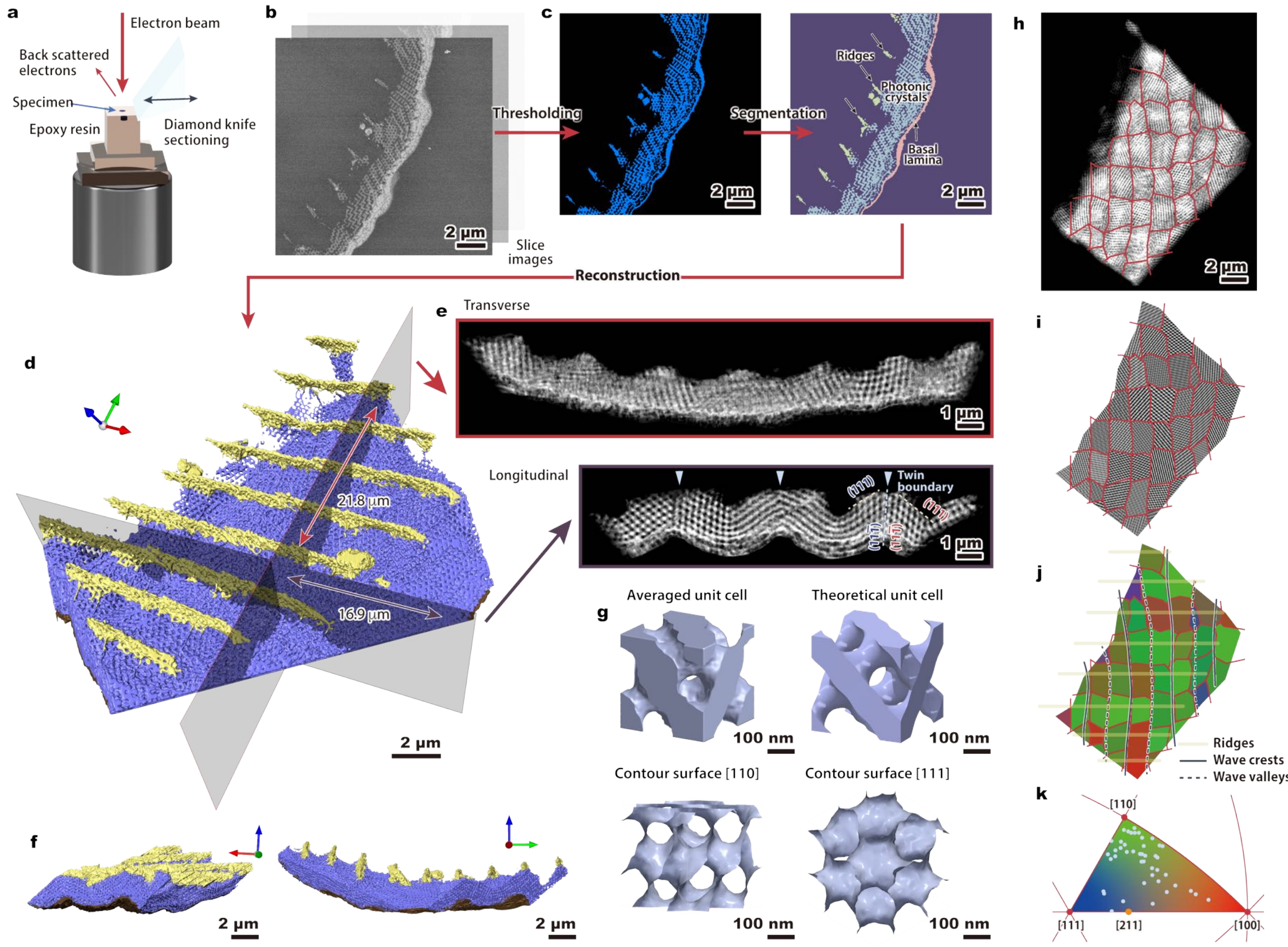

**Fig. 3. SBF-SEM 3D reconstruction of a dorsal scale. a-c,** Schematic representation of the cyclic processes and the image acquisition of the SBF-SEM data. Masks of the raw back-scattered electron images (**b**) were created by thresholding (**c**, left) then segmentation (**c**, right) to differentiate the components of the scale. **d,** Reconstructed 3D model, rotated to align with the sample axes; ridges, photonic structure, and basal lamina are colour-coded yellow, violet and brown, respectively. **e,** 0.6 µm slices of the model along the transverse (top) and longitudinal (bottom) directions displayed with volume rendering in additive blending mode. **f,** The model viewed from the transverse (left) and longitudinal axis (right), respectively. **g,** An average of 8 different unit cells of reconstructed PC and its contour surface view from [110] and [111], respectively, indicating a saddle-shaped surface closely resembling the theoretical diamond constant mean curvature family. **h,** Top view (optical axis) of the reconstructed model. Boundaries between crystal domains are illustrated in red. **i,** Theoretical projection of the SD domains along the experimentally determined orientations, generated with SPIRE[45]. **j,** Domain-orientation map of the reconstructed model, colour-coded. Positions of crests and valleys of the quasi-sinusoidal morphology and ridges are annotated. **k,** Partial inverse-pole figure showing the orientations of 44 domains.

2f). Analysis of the extinction conditions and wallpaper symmetries unambiguously assigns the structure to space group $Fd\bar{3}m$, which is the signature symmetry of the SD PC (Supplementary Note 3 and Supplementary Figs. 8,9). This assignment is further corroborated by overlaying a theoretical SD model onto the images. To the best of our knowledge, this is the first observation of an SD PC in Lepidoptera order[19,41,42].

Interestingly, the SD structure adopts a quasi-sinusoidal undulation, as revealed by the mosaicked TEM image acquired sequentially along the longitudinal axis (Fig. 2g). The crystal domains track the oscillations of the basal lamina, the curvature is relatively shallow near the basal surface, whereas quasi-twinning boundaries frequently appear at the crests of high curvature. These adjacent domains reverse orientation and expose {111} facets towards the overall abwing surface (Fig. 2h). This arrangement is consistent with the architectures observed on the dorsal surface in the SEM images (Fig. 2c).

### 3D determination of the hierarchical arrangement of the photonic structure

We then resolve this hierarchical organisation using serial block-face SEM (SBF-SEM) 3D reconstruction. The scales were embedded in epoxy resin and serially sectioned at 30 nm thickness (Fig. 3a), imaged by back-scattered electron imaging stacks (Fig. 3b and Supplementary Fig. 10) and reconstructed into a 3D volume, showing the ridges running parallel across the scale (yellow), perpendicular to the quasi-sinusoidal undulated PC (violet-blue), and the underlying basal lamina (brown) (Fig. 3d,f and Supplementary Fig. 11a-h). An average of eight unit cells shows that the chitinous skeleton is topologically an SD network with tetrahedral coordination (Fig. 3g). The volume fraction is calculated to be ~50% based on voxel counts (Supplementary Note 4).

Two typical 0.6 µm thick orthogonal cross-sections excluding ridges and basal lamina were visualised with volume-rendering: a longitudinal section showing the quasi-sinusoidal undulations (Fig. 3e bottom and Supplementary Video 2) and a transverse section intersecting seven ribbon-like ridges (Fig. 3e top and Supplementary Video 3). In the transverse view, the structure is largely flat at the base, with seven localised elevations as the attachment points of the seven ridges. In the longitudinal cross-section, the domains with [110]-like contrast following a quasi-sinusoidal undulation, exposing inclined (111) sidewalls. Additionally, the crystalline orientations of the SD PC form a correlated quasi-periodic domain grid (Fig. 3h and Supplementary Video 4).

To fully describe this orientation correlation, the orientation of each crystal domain was identified (Supplementary Table 1 and Supplementary Fig. 11i, see methods for detail). Notably, the domain boundaries are frequently located at the crests and valleys in the longitudinal direction, and midway between adjacent ridges transversely (Fig. 3i-k). This spatial correlation of the superstructure is essential to the quasi-periodic domain grid, suggesting a directed self-assembly process, in which geometric boundary conditions guide local crystallographic ordering. One ⟨111⟩ axis, typically close to the [111] direction, is largely normal to the quasi-sinusoidal undulation. A second ⟨111⟩ axis, (e.g. [11$\bar{1}$]), lies in the lateral plane perpendicular to the optical axis, forming an angle of ~70.5° with [111]. The in-plane orientation of this axis remains only weakly constrained, occurring both close to the longitudinal direction and at finite rotations relative to it. Notably, the quasi-twinning is often found at the crests of the sinusoidal profile, whereby neighbouring domains are related by mirror symmetry. The localisation of both the domain boundaries and the quasi-twinning[43,44] at the undulation extrema suggests that curvature plays a key role in regulating domain growth and inter-domain compatibility during self-assembly. Moreover, the crystallographic orientation statistics show that the angles of individual crystal domains are not unique, which gives rise to an angular range for the exposed inclined {111} facets.

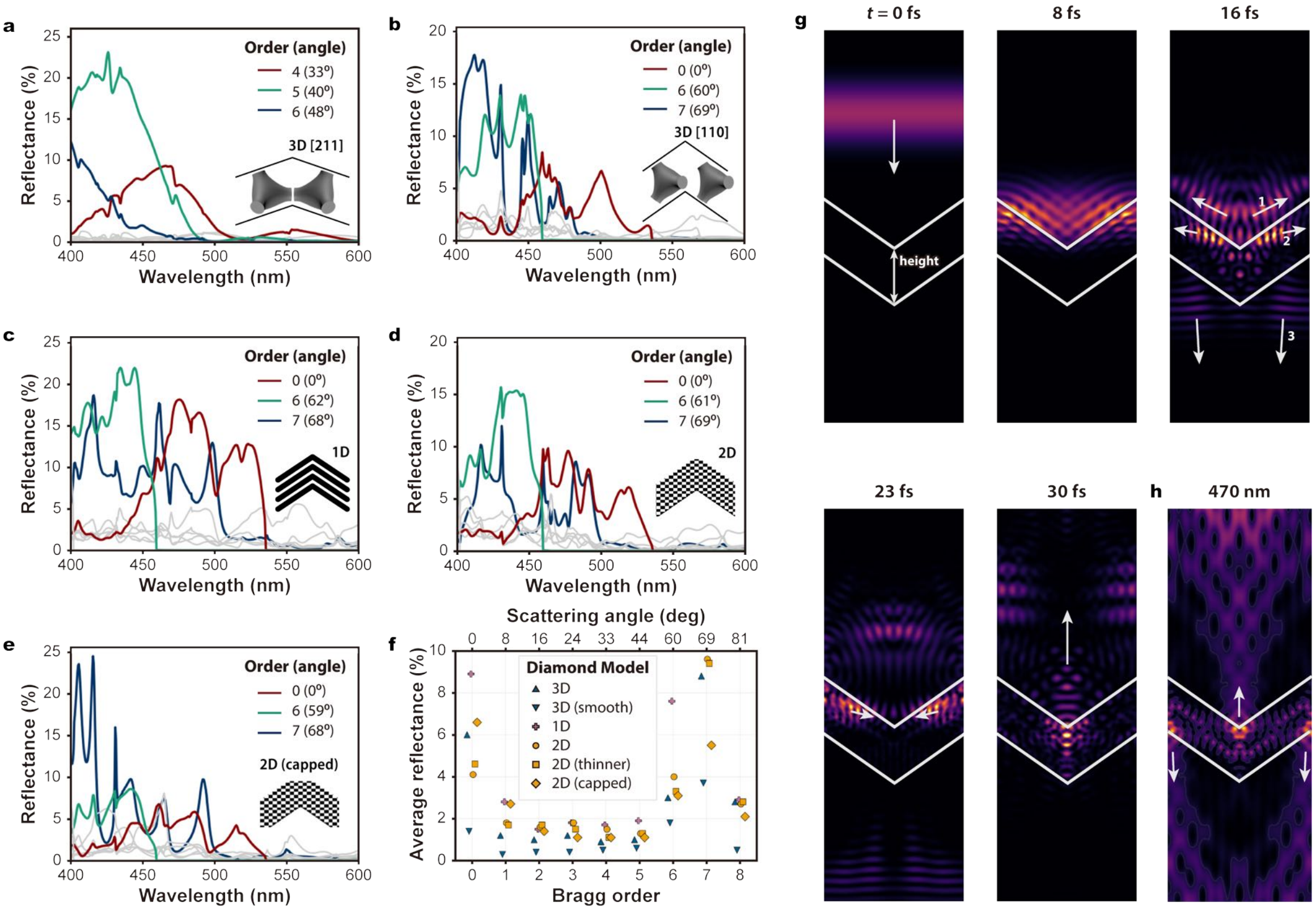

**Fig. 1. Optical analysis of the diamond structure in the Royal Jewel butterfly. a,** Twinned SD oriented with [211] along the slab-normal direction, producing zigzag facets with an inclination angle of ~20°. The dominant Bragg orders 4, 5, and 6 are highlighted. **b-e,** SD oriented with [110] along the slab-normal direction, resulting in facets inclined by ~35°. The corresponding models are: full 3D nodal surface (**b**); 1D Bragg-stack approximation (**c**); 2D model obtained by averaging along [001] (**d**); 2D model with capped zigzag canvas (**e**). The coloured curves indicate the dominant diffraction orders. **f,** Comparison of the different models, including a smoothly varying permittivity profile, 3D (smooth), based on the nodal function and a 2D model with reduced slab thickness, 2D (thinner). The top axis shows the reflectance-weighted mean scattering angle averaged across models. **g,** Time-resolved near-field response following normal-incidence excitation by a broadband Gaussian pulse. The incident field initially couples into the zigzag interface and subsequently separates into reflected, guided, and transmitted components. Repeated propagation through the periodic structure gives rise to a localised hotspot near the valley region. **h,** Representative QNM localised beneath the valleys. The mode radiates predominantly with near-vertical directionality, consistent with the observed angular redistribution.

## Structural origin of diffuse blue colour

To isolate the dominant scattering mechanism and preserve the underlying physical mechanism responsible for angular redistribution, we consider two representative crystal orientations observed in the reconstructed structure and replace the smooth undulations with a piecewise-linear zigzag profile (Fig. 3h-k and Supplementary Note 5):

1) A twinned SD with its [211] axis aligned to the vertical ($z$) axis. Twinning at the zigzag crest yields a [111] direction that is approximately perpendicular to the zigzag facets, corresponding to an elevation angle of ~20°.
2) A crystal with its [110] axis along $z$, for which two distinct ⟨111⟩ directions are normal to the zigzag facets, corresponding to an elevation angle of ~35°.

Since the zigzag period greatly exceeds the diamond lattice constant, the inclined {111} sidewalls act approximately as locally planar Bragg reflectors (Supplementary Fig. 12a). For case 1, this geometry redistributes blue light into medium scattering angles around 40°, corresponding to discrete Bragg orders in the idealised model and confirmed by the simulated spectrum in Fig. 4a. In the physical butterfly structure, disorder is expected to broaden this discrete response into a smooth angular distribution spanning approximately 30 to 50°.

For case 2, the inclined {111} sidewalls reflect blue light into higher scattering angles around 70°, as confirmed by the simulated spectrum in Fig. 4b. Unlike case 1), the response exhibits pronounced spectral modulations that depend sensitively on the crystal termination at the zigzag interface. Nevertheless, the overall angular redistribution remains robust. To suppress termination-specific variations, the spectra (Fig. 4b-e) are averaged over seven lateral offsets of the diamond structure within one zigzag period. Both orientations, therefore, exhibit a predominantly specular response, although the spectrum of case 2 contains a distinct resonant contribution in the 0$^{th}$ Bragg order that cannot be explained by Bragg reflection alone (see Supplementary Note 7). This indicates a second scattering pathway, in which incident light couples into resonant slab modes that subsequently radiate into the far field. The mechanism is qualitatively captured by a 1D Bragg-stack model (Fig. 4c), which reproduces the dominant scattering channels while simplifying the spectral response. A more realistic 2D reduction, obtained by averaging the SD nodal function along one crystallographic direction, yields a reflectance spectrum closer to that of the full 3D structure (Fig. 4d; see Supplementary Note 5 for model construction). While structural modifications affect the overall reflectivity and the distribution among individual Bragg orders, the angular redistribution remains remarkably robust across all models (Fig. 4e,f, Supplementary Notes 6-8).

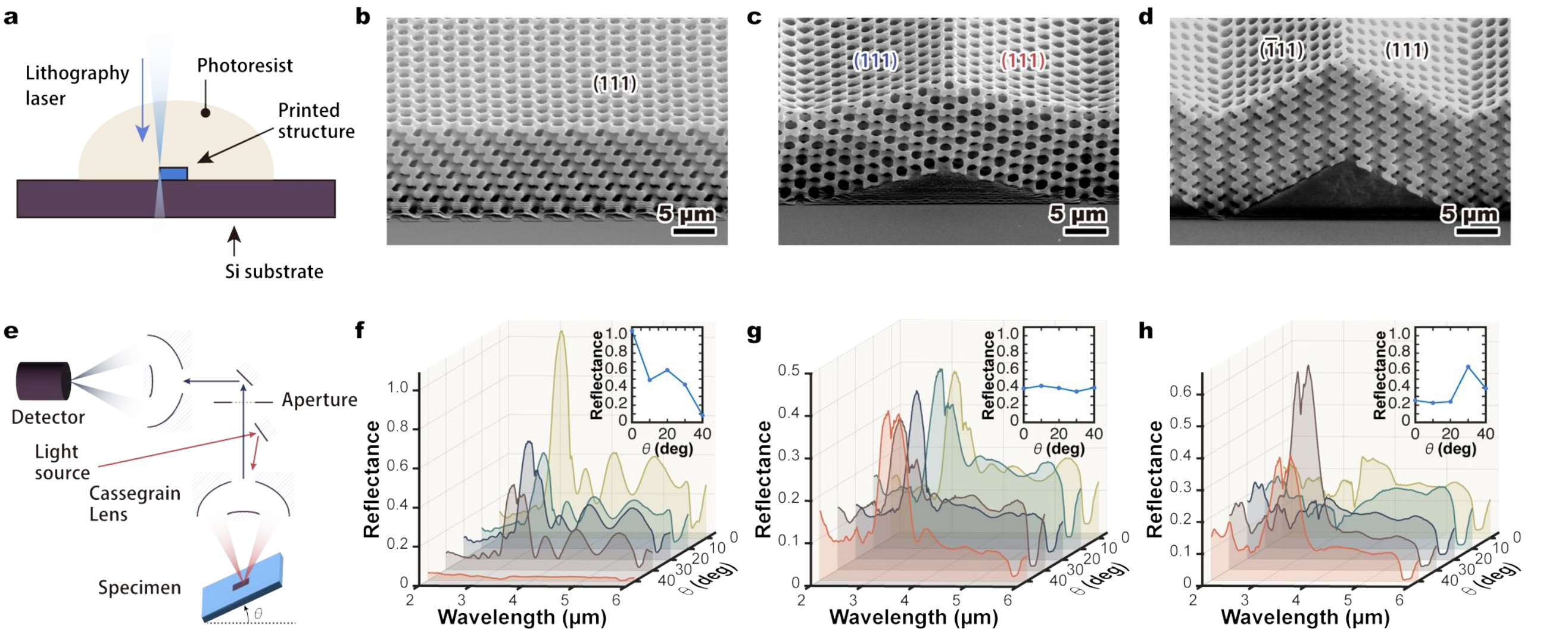

**Fig. 5. Artificial reconstruction of the undulating SD structure by TPP 3D printing. a,** Schematic of the TPP 3D printing workflow. **b-d,** SEM images of the printed structures. The structures are single crystal with [111] perpendicular to the substrate (**b**, Structure 1), twinned structure with [211] upwards (**c**, Structure 2) and single crystal with the [110] axis normal to the substrate (**d**, Structure 3), respectively. **e,** Diagram of the FT-IR spectroscopy, which focus on the specimen using a Cassegrain objective. **f-h,** Angle-resolved FT-IR reflectance spectra of the three printed structures. Insets display the angular dependence of the peak reflectance of the infrared bandgap. Panels **f**, **g** and **h** correspond to Structures 1, 2, and 3, respectively.

The scattering mechanism can be further understood through time-resolved simulations of a Gaussian pulse incident on the 1D Bragg-mirror model. As shown in Fig. 4g, the incoming light is initially partitioned between a strong specular reflection and weaker Bloch-like transport channels within the structure. These channels arise from strongly anisotropic propagating states whose group velocities are directed predominantly along the Bragg sidewalls (Supplementary Fig. 12c,d). Energy transported through these states is funnelled toward the valley regions, where it accumulates in localised hotspots. In the idealised periodic geometry, these hotspots are phase-matched and radiate predominantly into the 0$^{th}$ Bragg order. The out-coupling is mediated by quasi-normal modes (QNMs) localised beneath the valleys (Fig. 4h), which give rise to the resonant 0$^{th}$-order contribution observed in Fig. 4b-f. The corresponding resonances exhibit quality factors of approximately 50 and depend sensitively on the local zigzag geometry and crystal termination. The same physical picture persists in the more realistic 2D and 3D structures. Compared to the 1D model, these geometries support additional near-vertically propagating Bloch states (Supplementary Fig. 12b). Nevertheless, the density of these states remains small, particularly at shorter wavelengths, such that most light is still reflected and the valley-localised resonant channel remains active. Taken together, these results show that the angle-robust blue colouration originates from the combined action of specular reflection at inclined {111} facets and resonant scattering mediated by valley-localised QNMs.

## Towards biomimetic angular-insensitive reflectors: 3D-printed quasi-sinusoidal diamond structures

To explore the biomimetic potential of the hierarchical undulating diamond structure, three representative optical structures isolate the effects of topology, twinning and hierarchical modulation were artificially constructed through two-photon polymerisation (TPP, Fig. 5a), namely, a perfect single crystalline SD with a flat (111) surface (Structure 1), a twinned SD with [211] along *z*-axis (case 1 described above, herein Structure 2) and a zigzag-faceted single crystalline SD with [110] axis upward (case 2 described above, herein Structure 3).

Because of the instrument resolution limits, we scaled the unit cell to 3 μm, with 8 lattice planes along the ⟨111⟩ direction (Fig. 5a-d and Supplementary Fig. 13a-c). The printed micrometre-scale materials lack visible structural colour but reflect mid-infrared wavelengths. We thus employ Fourier transform infrared (FT-IR) microscopic reflection spectroscopy with specimen tilting (Fig. 5e and Supplementary Fig. 13d). In agreement with the expectation, Structure 1 shows a strong signal at 0° that falls to extinction at 40° (Fig. 5f). For Structure 2, the observed reflectance remains almost constant across the investigated tilt range (Fig. 5g). For Structure 3, the reflectance is subtle at lower angles but reaches its maximum at 30°, which is almost perpendicular to the exposed {111} surface (Fig. 5h). These results agree with the theoretical expectations, confirming that optical response is strongly governed by specialized crystallographic orientations and hierarchical morphologies. Only the finite low-angle response observed for Structure 3 is not reproduced by the simple geometric-optics treatment and may indicate an additional contribution from the internal transport and resonant scattering mechanisms discussed in the preceding section. Note that, the objective has a finite angular acceptance and therefore the measurements are qualitative rather than fully ARMS test (Supplementary Note 9 and Supplementary Fig. 14). Together with the successful fabrication of the printed structures, these results provide a proof-of-concept for biomimetic implementations of the quasi-sinusoidal diamond architecture.

## Discussion and conclusion

Lepidoptera (butterflies and moths) and Coleoptera (beetles and weevils) orders diverged approximately 250-230 million years ago[42,46-48]. As previously stated, the discovery of an SD photonic crystal in *H. polycletus* therefore substantially expands the known diversity of butterfly photonic architectures, which are otherwise dominated by SG morphologies[31,33,42]. This unexpected occurrence of an SD network in Lepidoptera suggests either convergent evolution towards a similar optical solution or a broader developmental accessibility of SD architectures than previously recognised.

The observed correlation between morphology, crystal orientation, domain placement and quasi-twinning suggests a growth mechanism for the SD PC under the geometric influence of the quasi-sinusoidal basal lamina. We hypothesize that folding of the cellular membranes first produces multiple nuclei beneath scale ridges. These nuclei develop into crystalline domains, while the sinusoidal chitin grooves of the basal lamina act as an architectural template that biases crystallographic orientation and guides domain growth. As domains expand, the quasi-sinusoidal geometry imposes spatially varying curvature. High-curvature regions (crests and valleys) act as barriers to further domain propagation because continued expansion increases geometric frustration. Consequently, domain boundaries localise at these positions, producing the quasi-periodic domain grid observed experimentally.

Importantly, the quasi-sinusoidal morphology is not curvature-symmetric; the basal lamina consists of broad valleys separated by relatively sharp crests. Because nucleation and early domain growth occur at this interface, domain boundaries are determined primarily by basal geometry. As the SD network develops, crystal planes follow approximately parallel surfaces, effectively reversing curvature asymmetry toward the top of the structure. Nevertheless, the domain configuration remains inherited from the original basal template. This asymmetry naturally explains the occurrence of quasi-twinning. Domains occupying neighbouring valleys eventually impinge at the crests, where mirror-symmetric configurations provide an energetically favourable means to accommodate the imposed geometrical constraints[43]. In contrast, twinning is rare in the weakly curved valleys, where reduced geometric frustration permits continued growth without the need for such accommodation. Curvature thus not only regulates domain placement but also governs inter-domain compatibility during self-assembly.

The hierarchical quasi-sinusoidal SD architecture produces a bright, angle-robust colouration distinct from previously reported biophotonic structures. The hierarchical modulation exposes inclined {111} facets that generate strong specular reflection through the fundamental ⟨111⟩ stopband. Domains whose {111} facets are inclined by approximately 20° relative to the wing surface normal predominantly scatter blue light into medium angles, whereas domains with more steeply inclined {111} facets (approximately 35°) preferentially scatter into both high and low angles. The high-angle response is specular, while the low-angle scattering is mediated by lateral Bloch transport and QNMs located at the valley bottoms. Our optical simulations and artificial prototypes confirm that the observed optical behaviour arises from the interplay between hierarchical modulation and the underlying SD photonic crystal, establishing a transferable design strategy for angle-robust photonic architectures.

Together, the Royal Jewel butterfly demonstrates a fundamental principle in nanophotonics: hierarchical modulation can substantially reshape the optical response of a photonic crystal under the refractive-index constraints imposed by biological materials. By coupling crystal topology, domain organisation, and morphology across different length scales, the quasi-sinusoidal SD architecture resolves the usual trade-off between the brightness of periodic photonic crystals and the angular robustness of disordered or polycrystalline designs. More broadly, this work reveals a previously unrecognised biophotonic design principle that links self-assembly, nanoscale architecture, and optical function. The resulting strategy suggests new routes toward both top-down and bottom-up fabrication of photonic materials, where hierarchical patterning can compensate for limited refractive-index contrast. The strong coupling between template geometry and SD organisation suggests a potential route toward template-guided self-assembly of complex photonic architectures, in which micron-scale templates direct the formation of nanoscale photonic order. This complements recent efforts in directed block-copolymer photonics[49], highlighting how nanoscale architecture can reshape optical functionality under severe material constraints and providing a general design principle for bioinspired nanophotonics.

## Methods

### Sample pretreatment

The typical male *H. polycletus* was commercially obtained from Indonesia and air-dried before investigation. Samples was taken from blue areas of the pristine wings using a surgery knife. For transmission electron microscopy (TEM) and serial block-face scanning electron microscopy (SBF-SEM) measurements, specimens were pretreated with $RuO_4$ straining. Typically, wings were firstly immersed into phosphate buffered saline (PBS) containing 2.5% (*w*/*w*) glutaraldehyde for 1 h, with a trace amount of sodium dodecyl sulphate (surfactant) added to avoid the anti-infiltration effect of the surface nanostructures. Then, the specimens were stained with 0.5% (*w*/*w*) $RuO_4$ aqueous solution for 1 h, followed by a gradient dehydration treatment with a graded series of ethanol/water mixtures (ending with 100% ethanol).

### Optical measurement

The specimens were directly viewed with a ZL500LPT polarised light microscope (Shanghai Fenye Optical Instrument Corporation) in reflection mode. The sample slide was mounted on a series of stages with different tilting angles to allow observing from multiple viewing directions. For tilted samples, images were taken using an extended depths-of-field technique to prevent image blurring.

The UV-Vis diffuse reflection spectra of the specific blue area of the wing were measured with an Agilent Carry 5000 UV-Vis-NIR spectrometer equipped with an integrating sphere. Prior to measurement the wing sample was trimmed to remove the black edges, leaving only the blue part. To confirm the refractive index matching effect, the sample was then infiltrated with acetophenone ($n = 1.5342$, 19 °C), the refractive index of which is close to that of chitin (1.55)[9]. After that the sample was rinsed with ethanol then leave the infiltrated liquid to evaporate. A spectrum was measured at each stage of the experiment. $BaSO_4$ powder was used as the white reference.

The angle-resolved reflectance spectra under diffuse illumination were measured with a microscopic angle-resolved micro-spectroscopy (ARMS) system based on an Ideaoptics NOVA high sensitive spectrometer. Measurements were performed on individual dorsal scale and the data of angle resolved reflectance spectra were averaged by 10 groups of parallel experiment to suppress noise. The diffused light source illuminated the specimen over incident angles of about ±60°, while the detector recorded from 0° to 90° with a step of 5°.

For the directional illumination, the spectra were measured using an Ideaoptics ARMS system. All the conditions matched those used for diffuse illumination measurements, except that a collimated incident light source was employed. The incident angles were swept from 0° to 50° in 10° increments, and for each incident angle the reflectance spectrum was measured continuously over a range of observation angles from −60° to +60°. Both angle-resolved spectra were measured using polytetrafluoroethylene (PTFE) Lambertian material as reference.

Back-focal-plane images were measured using an Ideaoptics ARMS system. Area of 40 μm in diameter from the specimen was selected with an aperture. The imaging used a 100× objective with numerical aperture (NA) of 0.9. Specimen was illuminated with a halogen lamp. For blue light illumination, wavelength of the light was 465 nm after inserting a filter.

Colours displayed according to angle-resolved spectra $R(\lambda)$ were simulated by computing their CIE tristimulus values and sRGB values. The calculation based on CIE 1931 2° standard observer colour matching functions $\bar{x}(\lambda)$, $\bar{y}(\lambda)$ and $\bar{z}(\lambda)$[50], preformed using MATLAB (R2025a, MathWorks). Tristimulus values $X$, $Y$ and $Z$ were calculated by integration with a 5 nm step ($\Delta\lambda$). Wavelength outside spectra range were ignored.

$$
\begin{aligned}
X &= k\sum_{\lambda=380}^{780} S(\lambda)R(\lambda)\bar{x}(\lambda)\Delta\lambda \\
Y &= k\sum_{\lambda=380}^{780} S(\lambda)R(\lambda)\bar{y}(\lambda)\Delta\lambda \\
Z &= k\sum_{\lambda=380}^{780} S(\lambda)R(\lambda)\bar{z}(\lambda)\Delta\lambda
\end{aligned}
\tag{1}
$$

Here, $k = 100/\sum_{\lambda} S(\lambda)\bar{y}(\lambda)\Delta\lambda$ serving as normalising factor, $S(\lambda)$ is relative spectral power distribution which represents average daylight. The results can be transformed to linear RGB using a standard sRGB (D65) conversion matrix[51,52], and then convert into non-linear sRGB via gamma correction:

$$
\begin{bmatrix} R_{\text{lin}} \\ G_{\text{lin}} \\ B_{\text{lin}} \end{bmatrix} = \begin{bmatrix} 3.2406 & -1.5372 & -0.4986 \\ -0.9689 & 1.8758 & 0.0415 \\ 0.0557 & -0.2040 & 1.0570 \end{bmatrix} \begin{bmatrix} X/100 \\ Y/100 \\ Z/100 \end{bmatrix} \tag{2}
$$

$$
C_{\text{sRGB}} = \begin{cases} 12.92C_{\text{lin}}, & C_{\text{lin}} \le 0.0031308 \\ 1.055C_{\text{lin}}^{1/2.4} - 0.055, & C_{\text{lin}} \ge 0.0031308 \end{cases} \\ C \in \{R, G, B\} \tag{3}
$$

### Structural analyses of the butterfly wing scales

Scanning electron microscopy (SEM) were conducted on JEOL JSM-7900F electron microscope with acceleration voltage of 0.8-1.0 kV. TEM observations were performed using JEOL JEM-F200 microscopy equipped with a Schottky gun operated at 200 kV ($C_s$ 1.0 mm, $C_c$ 1.4 mm, point resolution 0.23 nm for TEM mode and 0.16 nm for STEM mode). Images were recorded with a GATAN Rio camera (4096×4096 pixels). For TEM analysis, the specimens were embedded in epoxy resin and sectioned to thickness of 200 nm and 500 nm using a microtome. For SBF-SEM analysis, $RuO_4$ stained scales were embedded in epoxy resin and serially sectioned into 30 nm slices using an ultramicrotome. Each exposed surface was imaged by a Zeiss Sigma 300 VP SEM at an acceleration voltage of 2 kV, a dwell time of 3 μs per pixel under low vacuum condition (19 Pa). Serial back-scattered electron images were collected to generated an image stack comprising 2048×2048×199 voxels (voxel size 7.4 nm × 7.4 nm × 30 nm). The dataset was resampled to 1024×1024×398 voxels (voxel size 14.8 nm × 14.8 nm × 15 nm) to reduce anisotropy. The volume was processed and rendered in three-dimensional (3D) space using Avizo software. The model was rotated so that the longitudinal axis aligned with the *x*-axis in the Cartesian coordinate system, the transverse axis with the *y*-axis, and the abwing direction with the *z*-axis. For further analysis, eight unit cells were cropped from the model and averaged to generate a mean unit cell. The averaging was performed in MATLAB (R2025a, MathWorks) by converting the segmented models into binary matrices, and calculating their algebraic mean.

To identify the crystallographic domains orientation, each domain in the reconstructed model was examined by tilting to identify the [110] axis, then rotate to orient the [001] axis upwards and [010] axis to the right. The coordinates of these axes in the Cartesian coordinate system were obtained from the camera parameters displayed in Tomviz software (https://tomviz.org). These parameters were then calculated to convert to the corresponding crystallographic orientations toward *z*-axis and *y*-axis. In order to convert the direction of zone axes to directions toward *z*- and *y*-axes, we considered coordinates of [110] and [001] directions to be $\mathbf{v}_{011} = \mathbf{e}_y^C + \mathbf{e}_z^C$ and $\mathbf{v}_{001} = \mathbf{e}_z^C$, respectively. $\mathbf{e}_x^C$, $\mathbf{e}_y^C$ and $\mathbf{e}_z^C$ are primitive vectors of the lattice, $\mathbf{e}_x^C = \mathbf{e}_y^C \times \mathbf{e}_z^C$. The direction of *z*-axis or *y*-axis can be then expressed as $\mathbf{k} = u\mathbf{e}_x^C + v\mathbf{e}_y^C + w\mathbf{e}_z^C$, which is a linear combination of the primitive vectors. The $[uvw]$ indice of certain direction can be thus calculated by solving linear equations (4).

$$M\begin{pmatrix}u\\v\\w\end{pmatrix}=\mathbf{k}$$
$$M=\left[\mathbf{e}_x^C,\mathbf{e}_y^C,\mathbf{e}_z^C\right] \quad (4)$$

Small-angle X-ray scattering (SAXS) data was collected from a wing sample at beamline BL10U1 of the Shanghai Synchrotron Radiation Facility (SSRF). The measurement employed 10 keV photons with a 5-second exposure time. The data were processed with *SGTools* software[53].

## Optical simulations

The optical response of the quasi-sinusoidal diamond structures was investigated using a combination of frequency-domain band-structure calculations, time-domain electromagnetic simulations and quasi-normal mode (QNM) analysis.

### Photonic band structures and isofrequency contours

Photonic band structures and isofrequency contours (IFCs) were calculated using the MIT Photonic Bands (MPB) package. The dielectric structures were represented on a discrete spatial grid using level-set thresholding of the corresponding nodal functions, with a refractive index contrast of 1.55. Eigenfrequencies were computed for Bloch wavevectors sampled throughout the first Brillouin zone. IFCs were obtained by interpolating the computed bands and extracting constant-frequency sections corresponding to selected free-space wavelengths.

For the one-dimensional Bragg-stack approximation, IFCs were obtained semi-analytically from the exact Bloch dispersion relation of a periodic dielectric multilayer. At fixed frequency, the in-plane wavenumbers were numerically obtained for densely sampled Bloch wavenumbers in the first Brillouin zone. This approach was used to analyse the emergence of the characteristic double-trochoid topology in the vicinity of the first stop band.

### Time-domain simulations

Electromagnetic pulse propagation and reflection spectra were simulated using the finite-difference time-domain (FDTD) solver Tidy3D (Flexcompute Inc.). Both segmented dielectric structures derived from thresholded nodal functions and continuous dielectric representations with smoothly varying refractive index were investigated. In the segmented models, the dielectric interfaces were generated by level-set thresholding of the nodal functions, whereas in the continuous models the refractive index followed the underlying nodal field directly.

Broadband pulsed sources were used to illuminate the structures under investigation. Periodic boundary conditions were applied laterally ($x$-$y$), while perfectly matched layers (PMLs) were used in the propagation direction ($z$) to suppress artificial reflections. Reflection spectra were obtained from frequency-domain field monitors located above the structure and normalised to the source spectrum. Time-resolved field distributions were recorded in the $x$-$z$ plane to analyse the redistribution of optical energy within the quasi-sinusoidal geometry and to identify transport pathways associated with the hierarchical morphology.

### QNM analysis

QNM calculations were performed using COMSOL Multiphysics employing the finite-element method. The analysis was restricted to the one-dimensional Bragg-stack approximation of the zigzag sidewalls in order to isolate the resonant scattering channels associated with the local Bragg geometry. The computational domains were terminated by perfectly matched layers to represent open boundary conditions. Complex eigenfrequencies and field distributions were calculated in the absence of external excitation. The resulting QNMs were used to identify radiative resonances associated with the valley regions of the zigzag morphology and to establish the connection between the time-domain energy accumulation and the observed low-order scattering channels.

## Artificial reconstruction of the structures

The micro structures were prepared using a two-photon lithography system, Quantum X shape (Nanoscribe GmbH). The resin IP-Dip2 was firstly dripped on the polished silicon substrate, and the 63× print head was immersed in the resin to print. The STL files of all the structures were generated and processed using MATLAB and Blender software, then imported in the DeScribeX software and sliced by 0.1 μm for the accuracy of fabrication. After the lithography, the sample was immersed in the 1-methoxy-2-propyl acetate (PGMEA) for 8 minutes to develop the 3D-printed structures and clean the unpolymerised resin. Then, the sample was immersed in the isopropanol for 10 minutes to remove the PGMEA and unpolymerised resin further. The focused ion beam scanning electron microscopy (FIB-SEM), Thermo Scientific Scios 2 DualBeam was used to image the samples, with an accelerating voltage of 10.00 kV and beam current of 0.20 nA. Specimen were sputter coated with gold before imaging.

The reflection spectra were measured on the Fourier Transform infrared (FT-IR) spectroscopy (PerkinElmer Spotlight 200i). The infrared light from the source was focused by the Cassegrain and directed onto the sample with an angle range of 15° to 36°. And the reflected light was collected with a HgCdTe (MCT) detector. The spectra were acquired from a 100 μm × 100 μm region at the centre of the sample. To measure spectra with different incident angles, the sample was tilted to the desired angle using a tilting stage, while the acquiring area remained centre on the sample and constant in size. All measurements were normalised to the reflection of a gold mirror. The resin exhibits a weak absorption near 3.4 μm that overlaps the reflection band (Supplementary Fig. 13d); however, the reflections of all samples remain strong, and this absorption does not affect the measurements.

## Data availability

All data are available in the main text or/and the supplementary materials. Other information related to this study is available from the corresponding author upon reasonable requests.

## Author Contributions

L.H. conceived the idea and led the project. Y.L. performed basic SAXS, SEM, and TEM measurements. Y.L., Y.X., W.Z. and Y.C. conducted ARMS optical measurements. Y.L. and B.C. carried out SBF-SEM imaging and 3D reconstruction. M.S. performed optical simulations. S.H. and X.F. conducted 3D printing and FT-IR tests. X.W., C.C., and H.C. contributed to the analysis of structure-optical phenomena. M.S., Y.C., and L.H. validated the data and interpreted the results. Y.L. and Y.C. drafted the initial manuscript, and M.S., X.F., B.C., Y.C., and L.H. revised it with input from all authors. All authors read and approved the final manuscript.

## Acknowledgements

The authors thank Prof. Xinbin Cheng and Dr. Chao Feng of Tongji University for their assistance with the ARMS measurement. The authors thank the Shanghai Synchrotron Radiation Facility (SSRF) for providing BL10U1 beamline for collecting synchrotron SAXS data. This work was supported by the National Natural Science Foundation of China (grant Nos. 22425303, 22472058) and the Fundamental Research Funds for the Central Universities.

## Declaration of AI and AI-assisted technologies

During the preparation of this manuscript, the authors used

Microsoft Copilot, the localised Qwen3.6:35b, and Grammarly to correct grammar and spelling. After employing these tools, the authors reviewed the manuscript as needed and take full responsibility for its accuracy. All scientific content, interpretations, analyses, and conclusions were independently developed, verified, and approved by the authors.

### Competing interests

The authors declare no competing interest.